\documentclass[usenatbib]{mnras}
\usepackage{graphicx}
\usepackage{amssymb}
\usepackage{amsmath}
\usepackage{multicol}
\usepackage{natbib}

\newcommand{\nodata}{...}
\newcommand{\swift}{\emph{Swift}\ }

\title{Classifying AGN by X-ray Hardness Variability}
\author{Uria Peretz \& Ehud Behar
    \\
Department of Physics, Technion, Haifa 32000, Israel}

\begin{document}
    \maketitle
    \begin{abstract}
        The physics behind the dramatic and unpredictable X-ray variability of Active Galactic Nuclei (AGN) has eluded astronomers since it was discovered. We present an analysis of \swift XRT observations of 44 AGN with at least 20 \swift observations. We define HR-slope as the change of Hardness Ratio (HR) with luminosity ($L$). This slope is measured for all objects in order to:  1. Classify different AGN according to their HR-HR-slope relation  and 2. compare HR-$L/L_\mathrm{Edd}$ trends with those observed in X-ray binaries for the 27 AGN with well measured black hole masses. 
        We compare results using a count-based HR definition and an energy-based HR definition. 
        We observe a clear dichotomy between Seyferts and radio loud galaxies when considering count-based HR, which disappears when considering energy based HR.
        This, along with the fact no correlation is observed between HR parameters and radio loudness, implies radio loud and radio quiet AGN should not be discriminated by their HR behavior.
        We provide schematic physical models to explain the observed transition between energy defined HR states. 
        We find Seyferts populate the high, hard, phase 
        of the HR-$L/L_\mathrm{Edd}$ diagram as well as do three radio loud objects. Two LINERs populate the low, soft, phase part of this diagram. Finally, radio loud objects are concentrated around small positive HR-slopes, while Seyferts track the HR phase diagram which may provide clues to the geometry of the corona.
    \end{abstract}
	\section{Introduction}\label{sec:intro}
		The X-ray emitting corona of Active Galactic Nuclei (AGN) has been studied extensively, yet many questions remain. Open questions range from the heating mechanism that creates the $\sim10^9$ K source, through the geometry and location of this plasma in the near AGN environment, to the explanations of the hourly and daily variations in both X-ray flux and spectral shape.
		
		Several components are identified in the X-ray spectra of AGN, tying the spectral shape to physical phenomena. Most spectra have a dominant powerlaw, attributed to a corona reprocessing seed accretion disk photons into X-rays through comptonization. Two more  components determine the hardness of a spectrum; A soft excess below 1 keV whose origin is still debated \citep{Done07,DAmmando08,Boissay16}, and a hard component attributed to the X-rays reflecting off the disk which manifests primarily in a $\sim20-100$ keV Compton hump \citep{George91}. Only the tail is observed below 10 keV.
		
		Beyond these prevalent observed components in AGN, suppression of the soft band is observed, attributed to photoelectric obscuration. Examples are Seyfert 2s \citep{Kinkhabwala02} and transient obscurers in Seyfert 1s observed only recently in high resolution \citep{Kraemer05,Kaastra14,Mehdipour16}, but may not be rare \citep[][]{Markowitz14}.
		
		Many observations focus on high resolution spectroscopy in order to extract accurate measurements of the X-ray emission. A complementary method probing spectral variability is to use broadband spectra along with extensive monitoring in order to study property changes of the corona. The Hardness Ratio (HR), provides a quantitative description of the spectral shape:	
		\begin{align}
		HR=(H+S)/(H-S)\label{eq:hr}
		\end{align} 
		where $H$ and $S$ are the count rates of a given telescope in the defined hard ($H$) and soft ($S$) bands.
		Through analysis of the  HR along with the information on its variability, more insight can be garnered on the X-ray emitting mechanisms.
		
		In addition to illuminating the physics of the X-ray source, hardness surveys also probe the further environment of the AGN such as absorbers (\citet{Suchkov06} use HR to identify absorbed sources).
	
		\subsection{HR and BHXRBs}
		When discussing HR, it is interesting to see whether AGN cycle through the same phase states as those observed in Black Hole X-ray Binaries \citep{McHardy06}.        
        A comprehensive overview of BHXRBs by \citet{Remillard06} covers emission states of BHXRBs in Section 4.         
        One of the main results presented in \citet{Remillard06} is the BHXRB spectral cycle, transitioning through a soft, thermally dominated state, a steep powerlaw dominated state, and a hard powerlaw state. These are associated with the emergence and dissipation of jets, corroborated by simultaneous radio observations. Fig. 4 in \citet{Fender05} demonstrates this cycle in a HR phase diagram. Spectral hardening is observed with a steep rise in intensity associated with a jet, followed by a shock softening of the spectrum and finally the jet dissipates and intensity drops to the quiescent state. The HR is used to classify these states quantitatively and in a model-independent manner.      
              
        Both \citet{Remillard06} and \citet{Fender05} describe the soft part of the spectrum as a hot accretion disk, emitting thermally at 1 keV. This component is then shadowed by optically thick, hard, and non-thermal X-ray emission associated clearly with radio emission, the hallmark of a jet. Following the increasing intensity and the disappearance of the disk component, the hard part becomes more optically thin allowing the soft thermal component, the disk, to shine through. This happens while intensity remains maximal, exhibiting both non-thermal and disk components. \citet{Fender05} associate this state with a second, faster jet. In any case the powerlaw in this state is steep, similar to that observed in the soft state, perhaps indicative of a bright disk.
      
        \citet{Wu08} measure a break in the $L_X/L_\mathrm{Edd}$--powerlaw slope relation (where $L_X$ and $L_\mathrm{Edd}$ are the X-ray and Eddington luminosities) in 6 BHXRBs, pointing to a possible transition between a radiatively inefficient accretion flow to standard disk accretion. This gives a connection of the BHXRB observed spectral states to  physical models explaining these transitions, and associating the accretion behavior with that of the HR.
		
		\subsection{HR and AGN}				
		An important connection between BHXRBs and AGN has been identified by \citet{McHardy06} who finds a break in the power spectral density in both AGN and BHXRB. He ties this time-scale with the size of the accretion disk in both cases, which correlates with the black hole mass for stellar and AGN scales. 
		
		Searches for HR states in AGN analogous to BHXRB have been carried out for 20 years. \citet{McHardy99} measure softening of spectral slope with intensity in two AGN, MCG 6-30-15 and NGC 5506, drawing an analogue with BHXRB. Two more examples are \citet{Emmanoulopoulos12} which find a harder-when-brighter behavior for NGC 7213, and \citet{Mallick17} who find a softer-when-brighter behavior for Ark 120. Unlike BHXRBs, due to the longer timescale no single AGN can been observed to transition in a full HR cycle. 
		                
		\citet{Gu09} showed in a large compilation of low luminosity AGN that the powerlaw slope flattens with $L_\mathrm{Bol}/L_\mathrm{Edd}$ between AGN  ($L_\mathrm{Bol}$ is the bolometric luminosity). The authors associate this harder-when-brighter behavior with the hard phase of the BHXRB phase diagram. Interestingly, when looking at the much more luminous PG quasars with $L/L_\mathrm{Edd}>0.1$, \citet{Shemmer06} observe a reversal, i.e. softer-when-brighter behavior, again by comparing different AGN. This change of behavior  is further observed between the luminous sample of \citet{Shemmer08} and the low luminosity LINERS presented in \citet{Younes11}. When going into the highest energy regimes, TeV Blazars have been seen in a few works to conform to a harder-when-brighter  behavior \citep{Brinkmann03,Ravisio04,Pandey17,Pandey18}.
		
		Recently a comprehensive study by \citet{Connolly16} measured the relation between the HR behavior of the X-ray spectrum and the intensity of the AGN in a Palomar selected sample of 24 AGN observed using \swift XRT, defining the soft band up to 2 keV, and the hard above. They find that primarily the selected AGN display a harder-when-brighter behavior, with only 6 showing no correlation or a softening with luminosity, though this may be due to varying absorption. The authors find that low luminosity Seyferts belong in the luminous hard or intermediate part of the BHXRB phase diagram. This is in line with the model of \citet{Falcke04}, who attempt to unify the BHXRB and AGN picture by suggesting low luminosity AGN as the hard state counterpart of the BHXRB, with the spectrum dominated by a non-thermal powerlaw component. 

        \subsection{HR Caveats}
		One has to be careful when defining the bands for HR. For example, \citet{Rani} performed a recent HR study of a few AGN, BL LACs and Seyferts observed by NuStar. They define the soft band as 3-10 keV, and hard up to 79~keV, which is useful for measuring the effects of the Compton reflection bump. They find no  correlation of HR with flux.
		
		Another example is \citet{Sobolewska09}, who fit powerlaw slopes to the AGN hard ($>$ 2 keV) band observed by RXTE.
		They find a softer slope with increasing luminosity	unlike the \swift sample of \citet{Connolly16}.
		The difference can be explained by the different choice of bands, where \citet{Sobolewska09} are mostly sensitive to changes in the Compton reflection bump, and they do measure large reflection factors on average.		
		\\
		
        The main goal in this paper is to characterize AGN through their spectral states. This is done using the HR and its dependence on luminosity in order to identify physical explanations for these states and their transitions. 
        In addition, we consider the AGN sample as a tracer of a single state cycle, and compare it with a single BHXRB cycle.
        
    \section{Sample and data reduction}\label{sec:selection}
    Our main goal is to analyze the change of HR with flux. 
    In this work we define Hard ($H$) as 2-10 keV and soft ($S$) as 0.4-2 keV in the rest frame. Usually the soft band is defined on \swift XRT observations starting from 0.3 keV. Here we use 0.4 keV to allow for objects  $z\lesssim0.3$ to be analyzed consistently.
      
    We reduce \swift XRT observations\footnote{https://swift.gsfc.nasa.gov/archive/} in the PC mode using \textit{xselect} within the heasoft\footnote{https://heasarc.gsfc.nasa.gov/lheasoft/} software package. We extract the source spectrum using a 20 pixel circle around the coordinates taken from Simbad\footnote{http://simbad.u-strasbg.fr/simbad/} \citep{simbad}, and background using an annulus up to 30 pixels.
    
    The selection criteria are:
    \begin{enumerate}
        \item Objects are part of the Veron AGN catalog \citep{Veron10}.
        \item At least 20 observations in \swift XRT PC mode have:
        \begin{enumerate}
            \item more than 500 source photons
            \item at least one pixel with more than 3 photons
        \end{enumerate}
    \end{enumerate}
    
    Criterion 2(a) ensures that the hard and soft part of the band contain more than 10 photons in all used observations, such that the Poisson uncertainty of each band in each observation is at most 30\%, usually much better. 
    Criterion 2(b) eliminates high background and smeared observations.
    Of these we drop 8 objects with $z>0.3$ (next smallest is 0.36), and are left with 44 objects as detailed in Table \ref{table:sample}.       
        \begin{table*}
            \centering
            \caption{Sample statistics}\label{table:sample}
            \begin{tabular}{|l|c|c|c|c|c|l@{$\pm$}l|c|c|}
				Object       & Type & \swift\textsuperscript{1} &$z$\textsuperscript{\ 2} & Distance\textsuperscript{\ 3} & $\log\mathrm{M/M_\odot}$& \multicolumn{2}{|c|}{$L_X\textsuperscript{5}$}
				& $L_X/L_\mathrm{Edd}$ & $L_R/L_X$\textsuperscript{6}  \\   
				& & Observations           &                     &   Mpc                        & & \multicolumn{2}{|c|}{$10^{44}$ erg s$^{-1}$}
				&&                       \\
				\hline
				&&&&&&\multicolumn{2}{|c|}{}&&\\[-0.3cm]                                                                                   
                1ES 0647+250   & BLL & 35  & 0.203 &  773 &  \nodata          &20     & 0.4               &\nodata  &   0.0002         \\
                1ES 0806+524   & BLL & 21  & 0.137 &  530 &  \nodata          &3      & 0.06              &\nodata  &    0.002         \\
                1ES 1959+650   & BLL & 20  & 0.047 &  185 &  \nodata          &3      & 0.05              &\nodata  &   0.0003         \\
                1ES 2344+514   & BLL & 76  & 0.044 &  174 &  \nodata          &0.6    & 0.02              &\nodata  &   0.0008         \\
                1H0323+342     & BLL & 93  & 0.063 &  247 &  \nodata          &0.7    & 0.02              &\nodata  &    0.002         \\
                3C120          & SY1 & 28  & 0.034 &  133 &    7.11$^{4,1}$   &0.6    & 0.01              &0.04     &     0.01         \\
                3C273          & BLA & 48  & 0.158 &  609 &    9.13$^{4,1}$   &20     & 0.3               &0.01     &     0.05         \\
                Akn 564        & SY1 & 27  & 0.025 &   99 &    6.42$^{4,6}$   &0.3    & 0.007             &0.09     & $5\times10^{-5}$ \\
                BLLAC          & BLL & 37  & 0.069 &  271 &  \nodata          &0.7    & 0.02              &\nodata  &             0.03 \\
                ESO 362-G18    & SY2 & 24  & 0.013 &   50 &    7.42$^{4,1}$   &0.04   & 0.001             &0.001    & $3\times10^{-5}$ \\
                H 1426+428     & BLA & 66  & 0.129 &  500 &    8.87$^{4,2}$   &7      & 0.2               &0.008    & $8\times10^{-5}$ \\
                IC4329A        & SY1 & 19  & 0.016 &   64 &    7.84$^{4,1}$   &0.3    & 0.006             &0.003    & $2\times10^{-5}$ \\
                I Zw 187       & BLL & 48  & 0.055 &  217 &  \nodata          &1      & 0.02              &\nodata  &           0.0006 \\
                Mrk 110        & SY1 & 20  & 0.036 &  140 &    6.64$^{4,1}$   &0.6    & 0.01              &0.1      & $1\times10^{-5}$ \\
                Mrk 1383       & SY1 & 21  & 0.087 &  340 &    8.67$^{4,1}$   &1      & 0.03              &0.002    & $9\times10^{-6}$ \\
                Mrk 180        & BLL & 20  & 0.046 &  181 &  \nodata          &0.6    & 0.01              &\nodata  &           0.0006 \\
                Mrk 335        & SY1 & 46  & 0.025 &  101 &    7.29$^{4,1}$   &0.1    & 0.004             &0.005    & $4\times10^{-5}$ \\
                Mrk 501        & BLL & 37  & 0.033 &  130 &  \nodata          &0.7    & 0.02              &\nodata  &            0.002 \\
                Mrk 509        & SY1 & 156 & 0.034 &  135 &    8.07$^{4,1}$   &0.6    & 0.02              &0.004    & $1\times10^{-5}$ \\
                Mrk 766        & SY1 & 65  & 0.013 &   50 &    6.17$^{4,1}$   &0.04   & 0.001             &0.02     &           0.0001 \\
                Mrk 817        & SY1 & 26  & 0.031 &  124 &    8.10$^{4,1}$   &0.2    & 0.008             &0.001    & $7\times10^{-5}$ \\
                Mrk 841        & SY1 & 42  & 0.036 &  144 &    7.81$^{4,1}$   &0.3    & 0.008             &0.003    & $3\times10^{-5}$ \\
                MCG-06-30-15   & SY1 & 223 & 0.008 &   30 &    5.82$^{4,3}$   &0.03   & 0.001             &0.03     & $4\times10^{-6}$ \\
                MR 2251-178    & SY1 & 22  & 0.064 &  252 &    8.44$^{4,1}$   &2      & 0.07              &0.006    & $2\times10^{-5}$ \\
                NGC 1275       & SY2 & 67  & 0.018 &   70 &    6.48$^{4,1}$   &0.2    & 0.004             &0.04     &             0.09 \\
                NGC 1365       & SY1 & 32  & 0.005 &   21 &  \nodata          &0.007  & 0.0003            &\nodata  &           0.0006 \\
                NGC 2617       & SY1 & 52  & 0.014 &   57 &    6.82$^{4,4}$   &0.07   & 0.003             &0.008    & $4\times10^{-5}$ \\
                NGC 3031 (M 81)& LIN & 93  & \nodata &  3.6 &    8.16$^{4,2}$ &0.0002 & $8\times10^{-6}$  &$10^{-6}$&            0.259 \\
                NGC 3227       & SY1 & 27  & 0.004 &   15 &    7.18$^{4,1}$   &0.005  & 0.0001            &0.0003   & $8\times10^{-5}$ \\
                NGC 3783       & SY1 & 19  & 0.010 &   38 &    7.29$^{4,1}$   &0.05   & 0.002             &0.002    & $3\times10^{-5}$ \\
                NGC 4151       & SY1 & 190 & 0.003 &   13 &    7.58$^{4,1}$   &0.008  & 0.0004            &0.0002   &           0.0001 \\
                NGC 4486 (M 87)& LIN & 32  & 0.004 &   16 &    9.82$^{4,7}$   &0.003  & $9\times10^{-5}$  &$10^{-6}$&              0.4 \\
                NGC 4593       & SY1 & 172 & 0.008 &   33 &    7.27$^{4,1}$   &0.03   & 0.001             &0.001    & $8\times10^{-6}$ \\
                NGC 5548       & SY1 & 62  & 0.016 &   64 &    8.03$^{4,1}$   &0.08   & 0.003             &0.0006   & $3\times10^{-5}$ \\
                NGC 6814       & SY1 & 61  & 0.005 &   20 &6.87$^{4,1}$       &0.01   & 0.0004            &0.001    & $8\times10^{-6}$ \\
                NGC 7469       & SY1 & 128 & 0.016 &   63 &    7.32$^{4,1}$   &0.09   & 0.003             &0.003    &           0.0004 \\
                PDS456         & SY1 & 19  & 0.184 &  704 &  \nodata          &4      & 0.1               &\nodata  &           0.0002 \\
                PG 1218+304    & BLL & 35  & 0.184 &  702 &  \nodata          &10     & 0.3               &\nodata  &           0.0003 \\
                PKS 0447-439   & BLL & 24  & 0.107 &  417 &  \nodata          &3      & 0.05              &\nodata  &            0.002 \\
                PKS 0548-322   & BLA & 43  & 0.069 &  271 &    8.15$^{4,5}$   &2      & 0.04              &0.001    &           0.0004 \\
                PKS 1424+240   & BLL & 24  & 0.160 &  615 &  \nodata          &7      & 0.2               &\nodata  &            0.003 \\
                PKS 2005-489   & BLL & 22  & 0.071 &  279 &  \nodata          &2      & 0.04              &\nodata  &            0.007 \\
                PKS 2155-304   & BLL & 46  & 0.116 &  451 &  \nodata          &9      & 0.1               &\nodata  &            0.001 \\
                S5 0716+71     & BLL & 67  & 0.300 & 1114 &  \nodata          &20     & 0.5               &\nodata  &             0.01 \\                
            \multicolumn{10}{l}{\scriptsize 1 Number of observations (defined as good, see Sec. \ref{sec:selection}).}\\
            \multicolumn{10}{l}{\scriptsize 2 \cite{simbad}, Simbad database}\\
            \multicolumn{10}{l}{\scriptsize 3 \citet{Wright06},http://www.astro.ucla.edu/~Ewright/CosmoCalc.html, James Schombert Python version.}\\
            \multicolumn{10}{l}{\scriptsize 4 \citet{Koss17}, 1: Table 9, H$\beta$; 2: Table 4: Velocity dispersion; 3: Table 9, H$\alpha$;} \\
            \multicolumn{10}{l}{\scriptsize \;\; 4: \citet{Fausnaugh17}; 5: \citet{Barth03} 6: \citet{Botte04} 7: \citet{Gebhardt11}} \\
            \multicolumn{10}{l}{\scriptsize 5 Mean observed X-ray luminosity corrected for galactic absorption in the 0.4-10 keV rest frame band. See sec \ref{sec:HRsum}. } \\
            \multicolumn{10}{l}{\scriptsize 6 X-ray--Radio loudness is 5 GHz flux (1.4 if 5 is missing) taken from NED (http://ned.ipac.caltech.edu/)} \\
            \multicolumn{10}{l}{\scriptsize \;\; over mean 2-10 keV flux of all observations.}\\            
            \end{tabular}
        \end{table*}

	    Some of these have had their HR analyzed before for example 5 objects; NGC 3227, NGC 4151, M 87, M 81, NGC 5548 have been studied by \citet{Connolly16}, and 3C273 has been examined in \citet{McHardy06}.
	    
        The AGNs in Table \ref{table:sample} are mostly Seyfert 1 and BL LAC objects with 3 Blazars, two LINERS, and two Seyfert 2s. These span orders of magnitude in distance, luminosity, $L/L_\mathrm{Edd}$, and X-ray radio loudness ($L_R/L_X$). 
        
        That all said, these objects do not constitute a statistically well defined sample since 
        they have been selected for extensive monitoring with \swift for different reasons. 
        These are objects that are interesting to the community, and some of their
        biases include; the brightest and/or nearest AGN, 
        perhaps selected through their radio activity, or thought to be highly variable. 
        
        Nonetheless, the wide range of  AGN parameters, their variability, along with the fact that the sample includes a significant population of both radio loud and radio quiet AGN, make it interesting  for analyzing  spectral variability in AGN.       
          
    \subsection{HR analysis}\label{sec:anal}
    \subsubsection{Initial look - $F_\mathrm{var}$}    
        \begin{figure}
            \includegraphics[width=\linewidth]{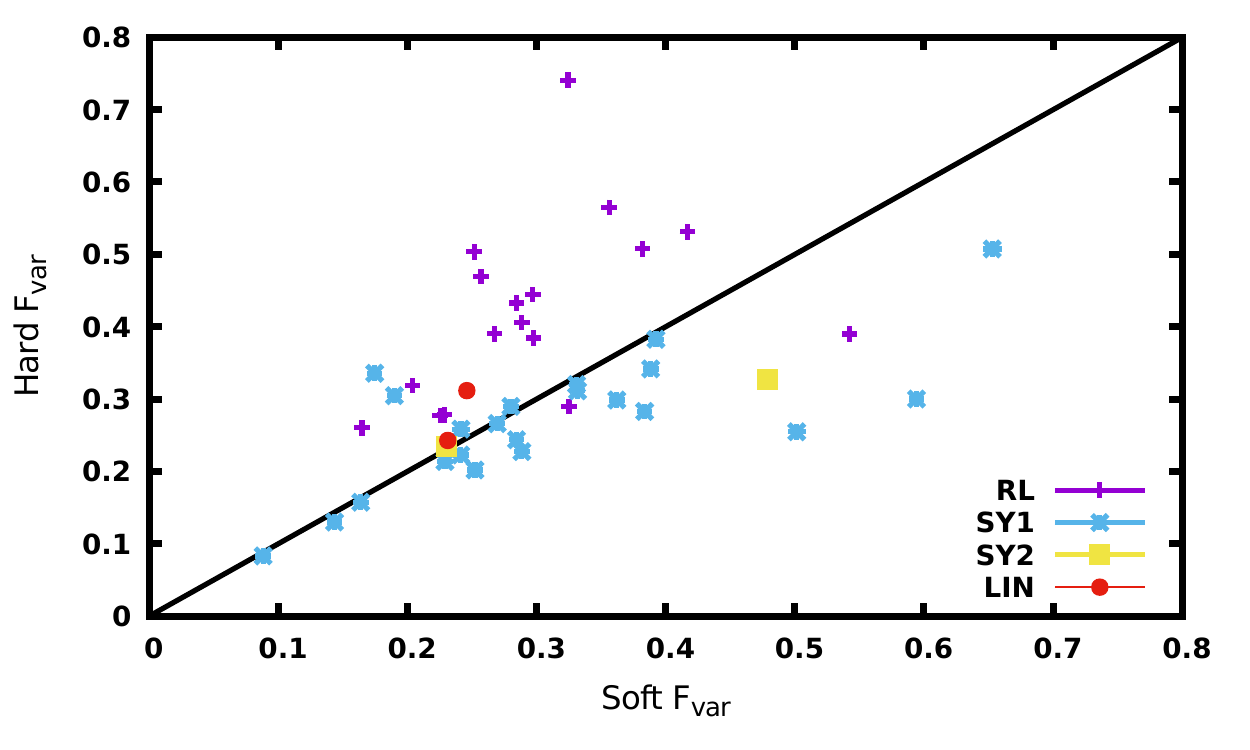}
            \caption{
            	Hard band variability plotted against Soft band variability. The equal variability line  is plotted for reference. 
            	Radio loud and LINERS are distinctly more variable in the hard band, while the Seyferts vary more in the soft band.
            }\label{fig:fvar}
        \end{figure}

        In the final column of Table \ref{table:hr} we present the excess variance parameter \citep{Markowitz4}:
        \begin{align}
        F_\mathrm{var}=\sqrt{\frac{\sigma^2-\overline{\sigma_i^2} } {\overline{C_i^2}} }
        \end{align}
        where the lightcurve variance is $\sigma^2$, 
        $\sigma_i^2$ are the count rate variances of each individual observation ($i$) representing the uncertainty, and
        $C_i$ are the individual observation count rates so that $\sigma^2=\overline{C_i^2}-\overline{C_i}^2$. An overline represents an average across all observations.      
        $F_\mathrm{var}$ is a measure of how much an object changes in excess of the observational uncertainty.
		The sample spans values of $F_\mathrm{var}$ from 7\% to 73\%.
        
        In Fig. \ref{fig:fvar} the soft band $F_\mathrm{var}$ is plotted against the hard, dividing AGN into two classes.
        Moreover, these objects are distinct to begin with, as AGN above the line are radio loud, and those below Seyferts.
        Different processes likely dictate variability for objects below and above the line,
        for example jet compared with outflow variability. This measurement already yields a separation through hardness behavior of these objects.
        Analysis using $F_\mathrm{var}$ is complementary to that presented in the following HR analysis, where we study variability of the hardness state with luminosity.            
    
    \subsubsection{Count HR}
    In contrast with the common definition of HR, in this work we use flux based definitions. Instead of considering count rates, in each channel (energy) we divide the count rate by the effective area. This allows for an instrument independent analysis that can be compared with past and future telescopes.
    After correcting all incoming photon energies for redshift, we sum count fluxes (counts s$^{-2}$ cm$^{-2}$) from 0.4 keV to 2 keV, and from 2 keV to 10 keV as emitted in the rest-frame
    and corrected for nominal galactic absorption \citep{Wilms00} in each channel individually. 
    The flux in each energy channel is calculated as:
    \begin{align}
    F_i^c=\frac{C_i}{T_i\cdot A_i}\label{eq:countflux}
    \end{align}
    where $C_i$ are the count rates in the channel, $T_i$ the galactic transmission at the channel energy as calculated by \emph{tbabs\footnote{http://pulsar.sternwarte.uni-erlangen.de/wilms/research/tbabs/}} \citep{Wilms00,Kalberla06} using abundances from \citet{Aspl09}, and $A_i$ is the effective area of the channel.
    The sums of these below and above 2 keV constitute the soft and hard bands used in eq. \ref{eq:hr}. Count uncertainties are propagated to the $H$ and $S$ band sums using Gaussian error propagation.
    We do this for each observation, and plot them against the mean count luminosity of the observations, as seen in the two top panels of Fig. \ref{fig:examples}, with the entire sample available on-line\footnote{Figure links}.
    
    We next use the Orthogonal Distance Regression (ODR) package \citep{Boggs89} as implemented in Python 3, Scipy\footnote{https://docs.scipy.org/doc/scipy/reference/odr.html} to get a best-fit HR-slope. 
    Using an ODR rather than a standard regression relaxes the assumption that HR is the dependent variable, taking into account more uncertainty.
    
    We employ the HR-slope to quantify the behavior of HR with luminosity, and ignore the intersection with the HR axis, as HR is not defined at zero luminosity. The hardness of an object is defined as the HR-mean across all observations. Details of the fit are listed in Table \ref{table:hr} in the 3rd and 4th columns. A positive slope indicates an harder-when-brighter behavior, and a negative one softer-when-brighter.    
    While a linear fit is not always a good one, it distinguishes between these two generic scenarios, and shows how drastically an object becomes harder-when-brighter or softer-when-brighter. Usually the trend is clear and the HR-slope is inconsistent with 0. In section \ref{sec:HRsum} we  discuss the results and these trends.
    
    \subsubsection{Energy HR}\label{sec:ehran}
    \begin{figure*}
    	\centering
    	\includegraphics[trim=0 0 3.5cm 2.5cm,clip,width=0.9\linewidth]{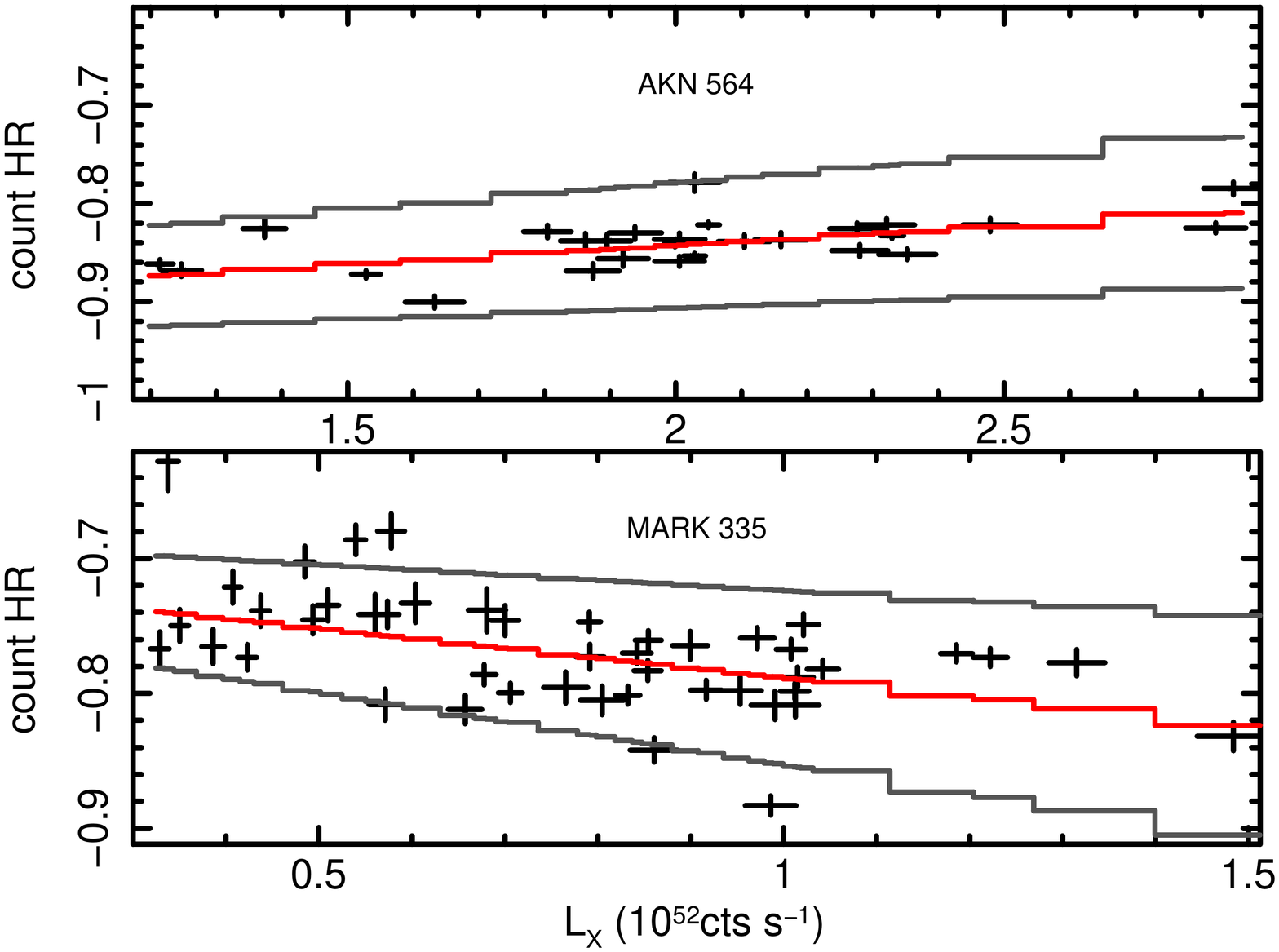}
    	\includegraphics[trim=0 0 3.5cm 2.5cm,clip,width=0.9\linewidth]{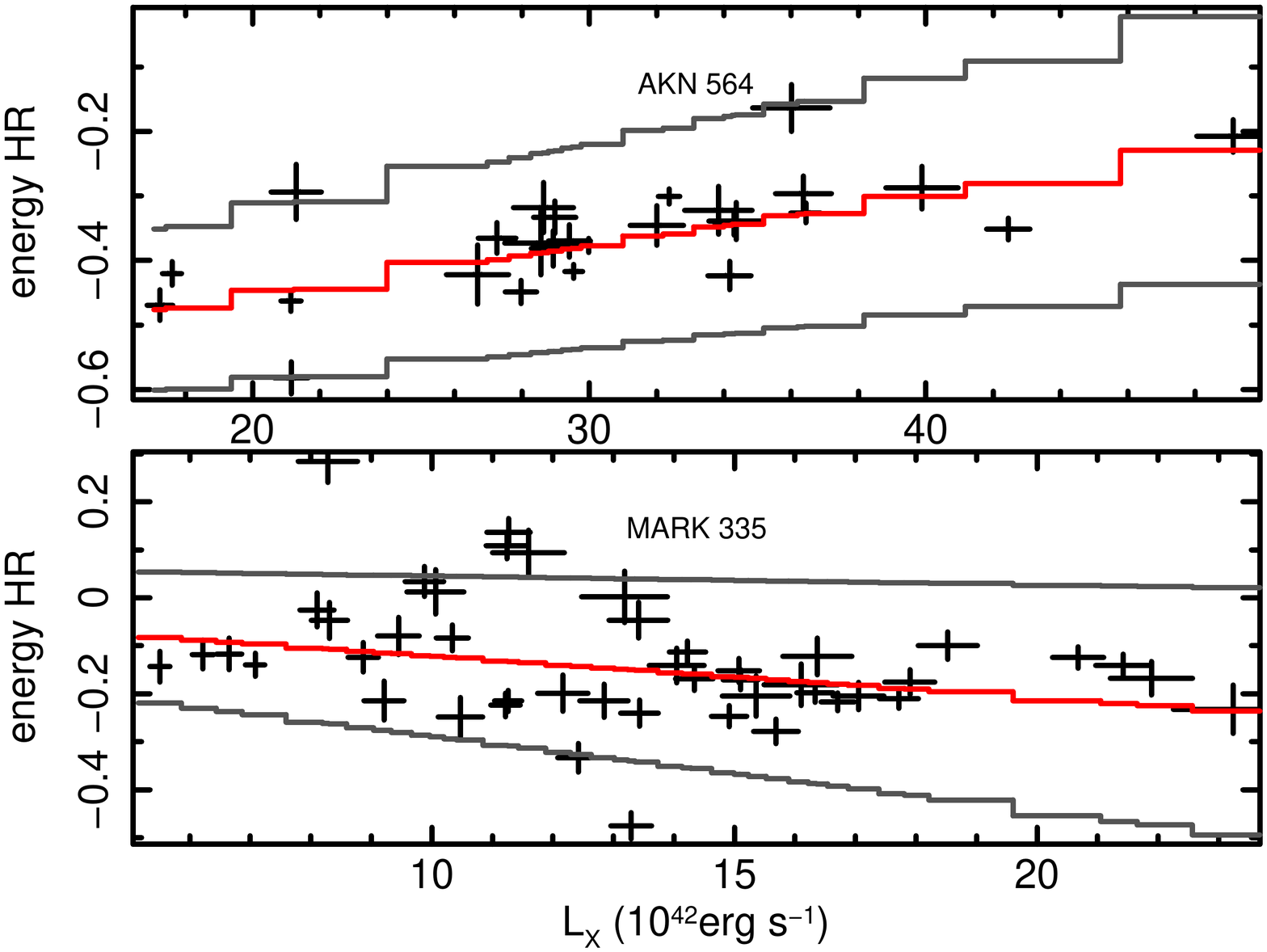}
    	\caption{Examples of harder-when-brighter (Akn~564, positive slope) and softer-when-brighter (Mrk 335, negative slope) behaviors discussed in Sec. \ref{sec:anal} and Table \ref{table:hr}. The red central line is the best fit, and the two grey lines represent the 90\% confidence interval. The top two panels show results when using a count based definition of the HR, and the bottom two show the same for an energy based definition.}\label{fig:examples}
    \end{figure*}
    Since the analysis is redshift corrected we can analyze the HR with a more physical definition of hard and soft. Instead of taking pure counts, we multiply for each energy bin the count flux in Eq. \ref{eq:countflux} by the rest-frame energy associated with the  channel $E_i$, and divide by the width of the energy bin $\Delta E_i$, to obtain the energy flux density:
    \begin{align}
    F_i^E=\frac{E_i}{\Delta E_i}F_i^c=\frac{C_iE_i}{T_i\cdot A_i\Delta E_i}\label{eq:energyflux}
    \end{align}
    This method, while uncommon, should give a more physical view of the HR behavior, as it represents the changes in the energy content of the AGN, which govern the interaction with its surrounding environment. Consequently, this definition has higher HR values compared to the count based definition, as more weight is given to the higher energies. Results are given in columns 5 and 6 of Table \ref{table:hr}. Two examples are presented in the two bottom panels of Fig. \ref{fig:examples}, and the complete sample is available on-line\footnote{Figure links}.
    
	\begin{table*}
        \centering
        \caption{HR and Energy HR results of the sample}\label{table:hr}
		\begin{tabular}{|l| l@{$\pm$}l|l@{$\pm$}l|l@{$\pm$}l|l@{$\pm$}l|l@{$\pm$}l|}
			Object & \multicolumn{2}{|c|}{HR Mean} & \multicolumn{2}{|c|}{HR Slope}  & \multicolumn{2}{|c|}{EHR Mean} & \multicolumn{2}{|c|}{EHR Slope}& \multicolumn{2}{|c|}{$F_\mathrm{var}$\textsuperscript{1}}\\
                   & \multicolumn{2}{|c|}{} & \multicolumn{2}{|c|}{$10^{-52}$ counts$^{-1}$ s} & \multicolumn{2}{|c|}{} & \multicolumn{2}{|c|}{$10^{-44}$ erg$^{-1}$ s}& \multicolumn{2}{|c|}{ }\\
            \hline
		    &\multicolumn{2}{|c|}{}&\multicolumn{2}{|c|}{}&\multicolumn{2}{|c|}{}& \multicolumn{2}{|c|}{}& \multicolumn{2}{|c|}{ }\\[-0.3cm]
            1ES 0647+250   & -0.8 &        0.02 &      0.0005 &      0.0003    & -0.2 & 0.04 & $8\times10^{-5}$ &$3\times10^{-5}$ & 0.315 & 0.003\\
            1ES 0806+524   & -0.8 &        0.01 &       0.002 &       0.002    & -0.4 & 0.04 &           0.0004 &          0.0003 & 0.277 & 0.005\\
            1ES 1959+650   & -0.6 &        0.02 &        0.01 &       0.008    &  0.1 & 0.03 &            0.001 &          0.0003 & 0.230 & 0.003\\
            1ES 2344+514   & -0.6 &        0.03 &        0.04 &        0.01    & 0.05 & 0.06 &            0.004 &          0.0007 & 0.333 & 0.003\\
            1H0323+342     & -0.7 &        0.03 &       -0.02 &       0.007    & 0.03 & 0.05 &          -0.0007 &          0.0008 & 0.306 & 0.002\\
            3C120          & -0.4 &        0.02 &       -0.05 &        0.05    &  0.3 & 0.03 &            0.001 &           0.002 & 0.155 & 0.002\\
            3C273          & -0.4 &        0.02 &      0.0006 &      0.0007    &  0.4 & 0.02 &  $6\times10^{-5}$&$2\times10^{-5}$ & 0.238 & 0.002\\
            Akn~564        & -0.8 &        0.01 &        0.04 &        0.02    & -0.4 & 0.05 &            0.008 &           0.003 & 0.199 & 0.003\\
            BLLAC          & -0.5 &        0.04 &       -0.03 &        0.02    &  0.2 & 0.05 &            0.002 &           0.001 & 0.436 & 0.004\\
            ESO 362-G18    & -0.6 &        0.03 &        -0.8 &         0.3    &  0.2 & 0.04 &            -0.07 &            0.03 & 0.442 & 0.003\\
            H 1426+428     & -0.6 &        0.02 &       0.007 &       0.002    & 0.05 & 0.04 &           0.0005 &$6\times10^{-5}$ & 0.178 & 0.002\\
            IC4329A        &  0.1 &        0.03 &        -0.2 &         0.4    &  0.7 & 0.02 &            0.005 &           0.003 & 0.128 & 0.003\\
            I Zw 187       & -0.7 &        0.02 &        0.03 &       0.006    &-0.07 & 0.05 &            0.002 &          0.0003 & 0.387 & 0.002\\
            Mrk 110        & -0.6 &        0.02 &       -0.03 &        0.04    &  0.2 & 0.03 &            0.003 &           0.003 & 0.082 & 0.003\\
            Mrk 1383       & -0.7 &        0.02 &      -0.004 &       0.005    &-0.09 & 0.06 &          -0.0002 &          0.0009 & 0.224 & 0.005\\
            Mrk 180        & -0.8 &        0.01 &        0.07 &        0.02    & -0.2 & 0.04 &            0.005 &           0.001 & 0.374 & 0.003\\
            Mrk 335        & -0.8 &        0.02 &       -0.07 &        0.03    & -0.1 & 0.06 &           -0.009 &           0.007 & 0.349 & 0.003\\
            Mrk 501        & -0.6 &        0.02 &        0.07 &        0.03    & 0.08 & 0.05 &            0.005 &           0.001 & 0.288 & 0.003\\
            Mrk 509        & -0.6 &        0.03 &       -0.04 &       0.009    &  0.2 & 0.05 & $7\times10^{-5}$ &          0.0008 & 0.269 & 0.002\\
            Mrk 766        & -0.7 &        0.02 &       -0.06 &        0.08    & 0.05 & 0.06 &           -0.006 &            0.01 & 0.388 & 0.002\\
            Mrk 817        & -0.7 &        0.03 &        -0.1 &        0.06    &-0.05 & 0.07 &            -0.02 &           0.009 & 0.276 & 0.005\\
            Mrk 841        & -0.7 &        0.02 &       -0.05 &        0.03    & 0.04 & 0.06 &           -0.001 &           0.003 & 0.325 & 0.004\\
            MCG-06-30-15   & -0.5 &        0.03 &        0.04 &         0.2    &  0.3 & 0.06 &             0.03 &            0.01 & 0.278 & 0.001\\
            MR 2251-178    & -0.4 &        0.04 &       -0.02 &        0.01    &  0.4 & 0.04 &          -0.0002 &          0.0006 & 0.231 & 0.005\\
            NGC 1275       & -0.5 &        0.03 &        -0.2 &        0.09    &  0.3 & 0.04 &            0.002 &           0.004 & 0.224 & 0.002\\
            NGC 1365       &  0.3 &        0.04 &          -8 &           5    &  0.8 & 0.02 &             -0.1 &            0.06 & 0.543 & 0.004\\
            NGC 2617       & -0.6 &        0.03 &        -0.1 &        0.09    &  0.2 & 0.06 &            0.005 &           0.006 & 0.324 & 0.003\\
            NGC 3031 (M 81)& -0.6 &        0.04 &          80 &          30    &  0.2 & 0.07 &               10 &               2 & 0.259 & 0.003\\
            NGC 3227       & -0.2 &        0.03 &         -10 &           8    &  0.5 & 0.03 &             -0.7 &             0.3 & 0.330 & 0.003\\
            NGC 3783       & -0.3 &        0.05 &          -2 &         0.5    &  0.5 & 0.04 &            -0.08 &            0.03 & 0.379 & 0.005\\
            NGC 4151       &  0.5 &        0.05 &          40 &           4    &  0.9 & 0.01 &             0.07 &           0.006 & 0.268 & 0.002\\
            NGC 4486 (M 87)& -0.7 &        0.03 &        -0.4 &           2    & -0.2 & 0.06 &              0.3 &             0.2 & 0.229 & 0.004\\
            NGC 4593       & -0.5 &        0.04 &        -0.6 &         0.2    &  0.3 & 0.06 &             0.03 &            0.01 & 0.231 & 0.002\\
            NGC 5548       &  0.2 &        0.05 &          -4 &         0.7    &  0.7 & 0.03 &            -0.03 &            0.01 & 0.381 & 0.004\\
            NGC 6814       & -0.5 &        0.04 &          -2 &           1    &  0.3 & 0.06 &            -0.03 &            0.06 &  0.364& 0.003\\
            NGC 7469       & -0.6 &        0.03 &       -0.06 &        0.06    &  0.1 & 0.06 &             0.01 &           0.005 & 0.263 & 0.002\\
            PDS 456        & -0.7 &        0.04 &      -0.009 &       0.004    &-0.09 & 0.06 &          -0.0003 &          0.0006 & 0.226 & 0.006\\
            PG 1218+304    & -0.7 &        0.02 &       0.003 &      0.0008    &-0.06 & 0.05 &           0.0002 &$5\times10^{-5}$ & 0.290 & 0.004\\
            PKS 0447-439   & -0.8 &        0.01 &       0.004 &       0.002    & -0.4 & 0.04 &           0.0008 &          0.0004 & 0.290 & 0.004\\
            PKS 0548-322   & -0.6 &        0.02 &       0.007 &       0.005    &  0.2 & 0.04 &           0.0006 &          0.0003 & 0.234 & 0.002\\
            PKS 1424+240   & -0.8 &        0.02 &       0.002 &      0.0006    & -0.4 & 0.05 &           0.0003 &          0.0001 & 0.394 & 0.005\\
            PKS 2005-489   & -0.8 &        0.02 &        0.01 &       0.008    & -0.3 & 0.05 &            0.001 &          0.0006 & 0.643 & 0.004\\
            PKS 2155-304   & -0.8 &        0.01 &      0.0008 &      0.0006    & -0.3 & 0.04 &           0.0002 &$8\times10^{-5}$ & 0.296 & 0.002\\
            S50716+71      & -0.8 &        0.02 &     -0.0004 &$8\times10^{-5}$& -0.3 & 0.05 &-$9\times10^{-5}$ &$2\times10^{-5}$ & 0.518 & 0.004\\      
            
            \multicolumn{11}{l}{\scriptsize 1 Defined in Sec. \ref{sec:selection}.}\\             
		\end{tabular}
	\end{table*}
 
	\section{HR behavior of the sample}\label{sec:HRsum}
	\begin{figure*}
		\includegraphics[trim=0 0 0.3cm 0,clip,width=0.48\linewidth]{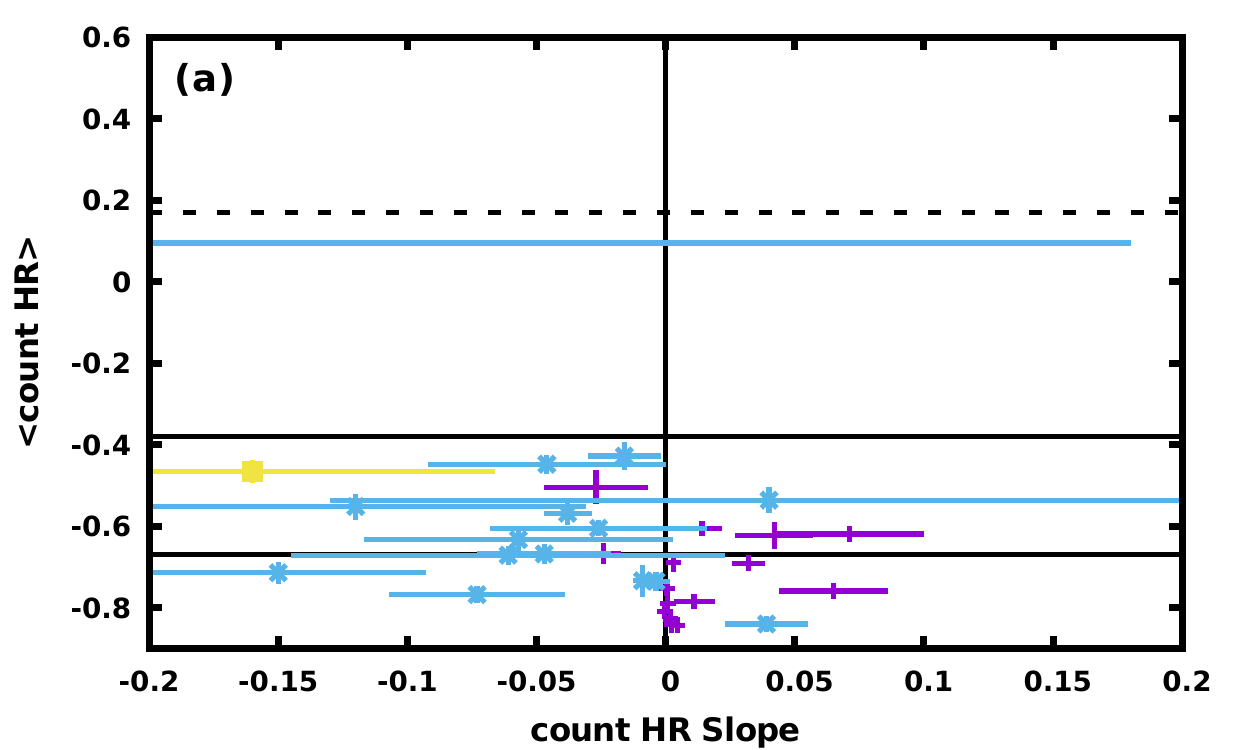}	
		\includegraphics[trim=0.5cm 0 0 0,clip,width=0.48\linewidth]{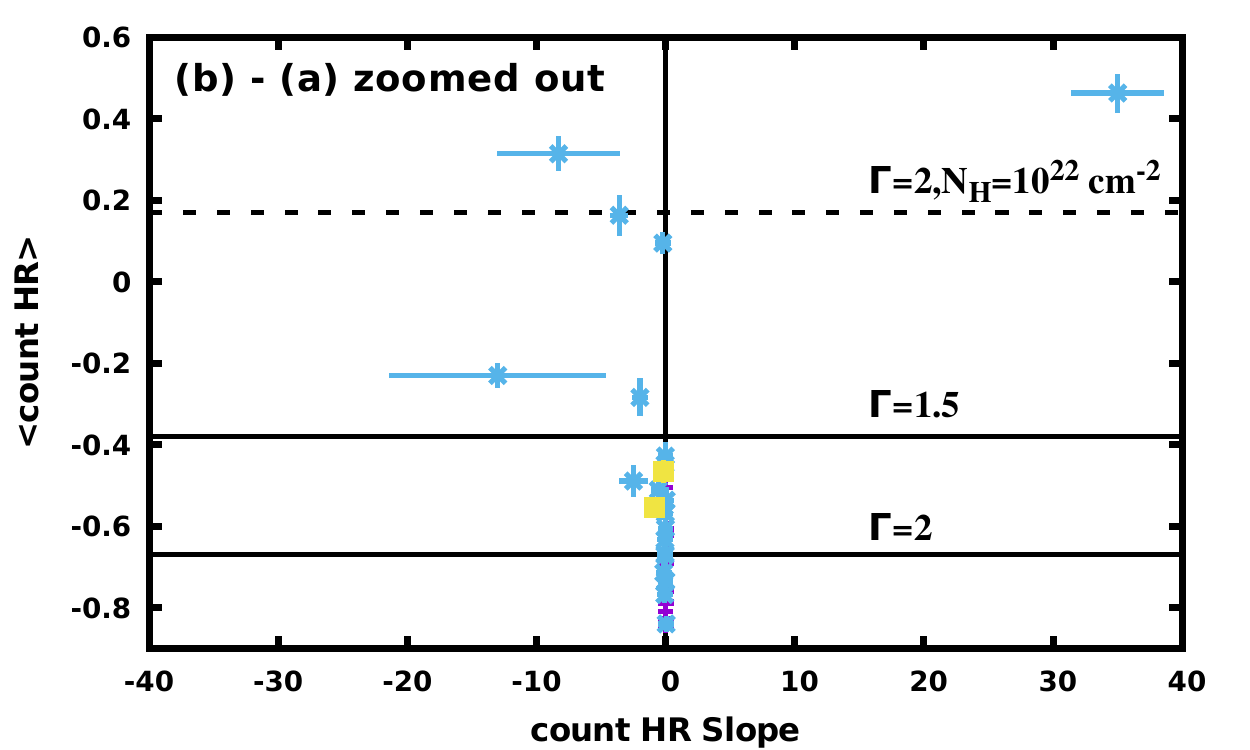}		
		\includegraphics[trim=0 0 0.3cm 0,clip,width=0.48\linewidth]{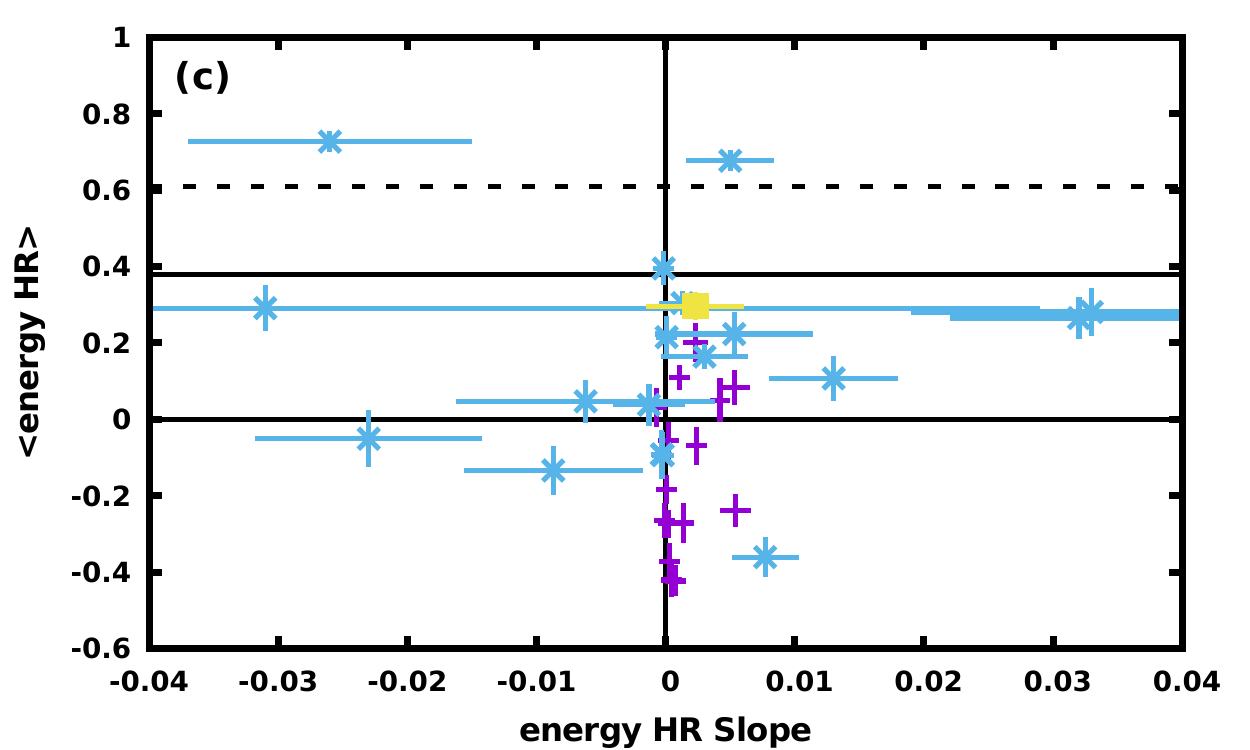}
		\includegraphics[trim=0.5cm 0 0 0,clip,width=0.48\linewidth]{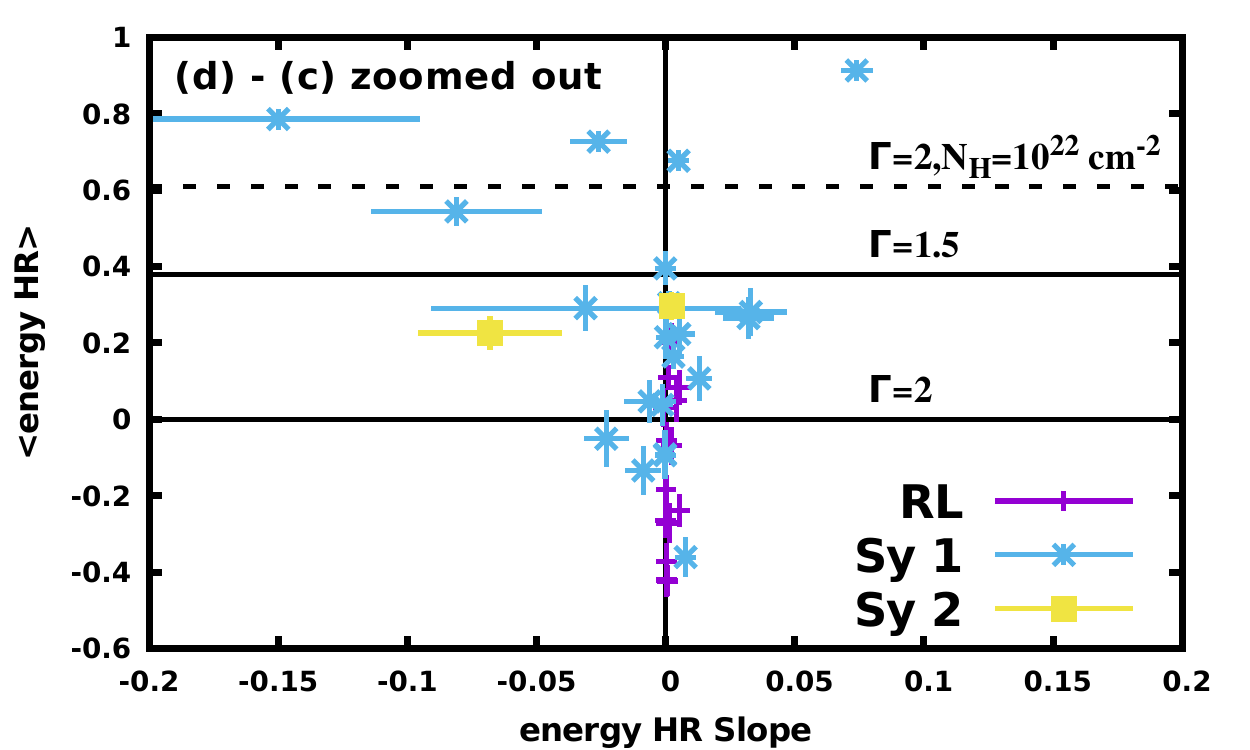}
		
		\caption{A graphical summary of results presented in Table \ref{table:hr}. Each plot is a HR-mean-slope phase diagram, two plots for each definition, count-HR 
			(\textbf{top}) or energy-HR (\textbf{bottom}). 
			The \textbf{left} plots are centered around the $\mathrm{HR}=0$ vertical line, and zoom-outs are on the \textbf{right}.
			(\textbf{right}). 
			Horizontal solid lines correspond to a HR of a $\Gamma=2$ and $\Gamma=1.5$ powerlaw. The dashed lines show a $\Gamma=2$ powerlaw absorbed by a neutral column of $10^{22}$ cm$^{-2}$. 
			When considering energy the mean hardness is increased, as a greater weight is given to the hard band.		
			The dashed line singles out heavily absorbed objects (above it). Most unabsorbed Seyferts are contained between the HR defined by the two powerlaw slopes, but when considering energetics (\textbf{bottom}) the predominantly softer-when-brighter behavior of the count analysis (\textbf{top}) is gone.}\label{fig:phase_diagrams}	
	\end{figure*}
	The summary of our main results is presented in Fig. \ref{fig:phase_diagrams}. These four plots show all objects on a HR-mean-slope  diagram, and are naturally divided by the 0 HR-slope line.
	On the right side of each diagram are harder-when-brighter objects, on the left softer-when-brighter. The bottom half are  softer objects, and the upper harder objects. Two plots are given for each definition of HR, counts (top) and energy (bottom) - one focusing on most objects (left), and a zoomed out plot showing the outliers as well (right). The two LINERS are omitted from these plots due to steep slopes with large error bars (see Table \ref{table:hr}).	
	
    The two solid horizontal lines are HR values of AGN spectral photon powerlaw indices of $\Gamma=1.5$ (top) and $\Gamma=2$ (bottom). The dashed line shows a $\Gamma=2$ powerlaw absorbed with a neutral column of $10^{22}$ cm$^{-2}$. Most AGN are contained as expected between the two powerlaws, with obscured objects such as NGC 5548 and NGC 1365 near and above the dashed line (right plots). NGC 4151, in the top far right panel is a clear obscured outlier.
	The two top panels are dedicated to the count based definition of HR. This is the classic value used, and interestingly when considering all objects provides a clear dichotomy between Seyferts and radio loud AGN (top left).
	Radio loud AGN are harder-when-brighter and Seyferts are softer-when-brighter, though perhaps it is more sensible to call this behavior harder-when-fainter as Seyferts often have changing ionized absorbers that can attenuate the soft band count much more. 
	
	It may be hard to attribute a physical meaning to the count HR, so we consider an energy based HR (bottom panels in Fig. \ref{fig:phase_diagrams}) and  the picture changes somewhat. Seyferts, while maintaining the bottom left to top right orientation on the plot, are shifted both in slope and hardness compared to the count diagram. While a hardening of all objects is an obvious consequence of the new energy definition, a steeper HR-slope is not immediately implied. The radio load AGNs for example, remain centered around HR-slope=0, remaining harder-when brighter, though this may be a consequence of being near a HR-slope of 0. Note NGC 4151 is no longer as peculiar in the bottom right panel, now that the hard counts are given a greater weight in the HR.

\begin{figure}
		\includegraphics[width=\linewidth]{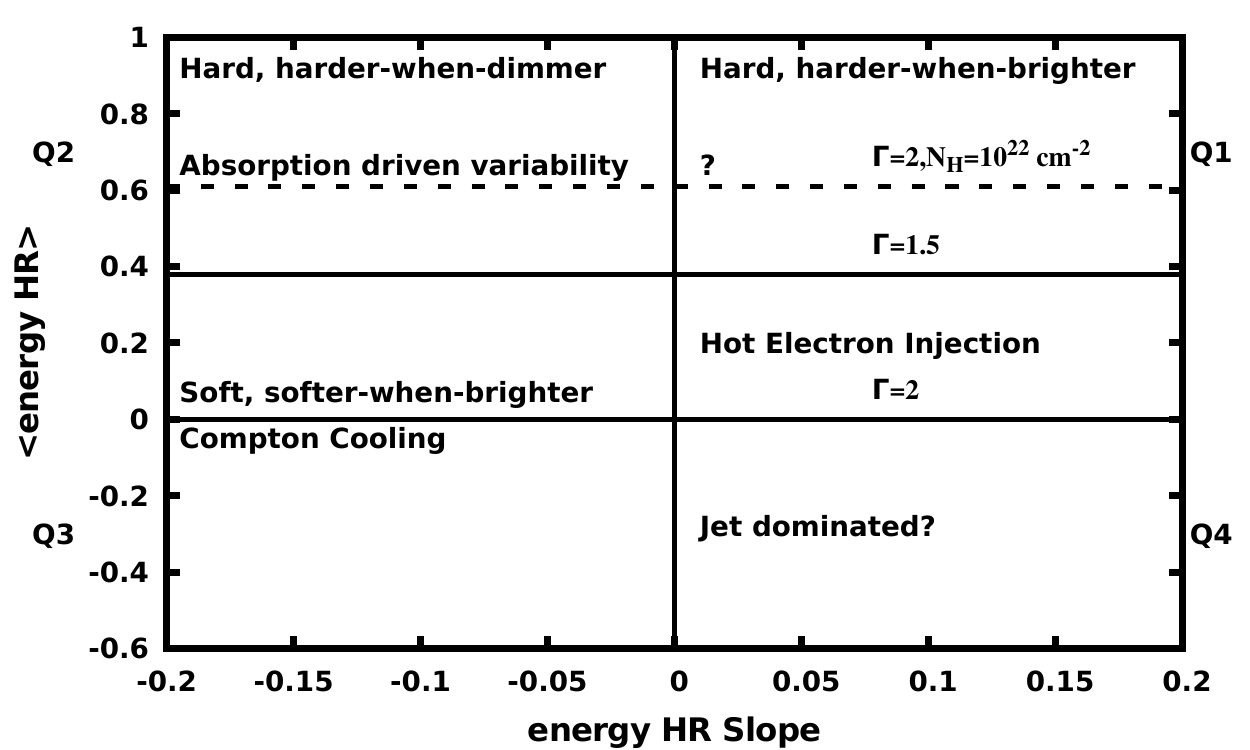}
		\caption{A qualitative description of the physical processes that may govern HR variability in each quadrant of the HR phase diagram.}\label{fig:emptyhr}
	\end{figure}

	The plot can be divided into 4 quadrants by the $\Gamma=1.5$ line with absorbed objects above it, perhaps even 6 with $\Gamma=2$ as a second horizontal axis. With the HR energy definition we can attribute clear physical meaning. Numbering 4 quadrants from top right counter-clockwise we have:
	\begin{enumerate}
		\item[Q-1] Harder-when-brighter objects that are initially hard to begin with. Any harder-when-brighter behavior can most easily be attributed to an injection of hot electrons into the X-ray emitting gas, whether through magnetic reconnection in the corona or through jets attributed to radio loud AGN. The Seyferts in this quadrant may comprise objects with the most active coronae in the sample. NGC 4151, the outlier in this quadrant, has a completely absorbed soft band, so any change in HR must be attributed to a change in the hard band. 
		\item[Q-2] Hard objects that become softer-when-brighter. Perhaps a better name is harder-when-fainter, as these objects are dominated by variable obscuration (NGC 5548, NGC 3783, NGC 1365).
		\item[Q-3] Objects with soft emission dominating and softer-when-brighter behavior. This quadrant could fit with the coronal cooling paradigm as detailed by \citet{Haardt91}, where a hot corona brightens due to increasing UV disk photons and softens.
		\item[Q-4] Soft objects with harder-when-brighter behavior. This quadrant is dominated by radio loud AGN, and may be attributed to a jet emerging and dominating a previously quiescent soft emitter, as in the BHXRB picture. Akn~564 is the only Seyfert in this quadrant, which is interesting as it is not associated with any special radio activity.
	\end{enumerate}	
	These 4 quadrants are depicted in Fig. \ref{fig:emptyhr}.
	\\
	
	The very different emission mechanisms interpreted from the full spectral energy distribution of the radio loud and radio quiet AGN, along with the fact the two mix in this diagram, suggests that the two types of objects should not be unified by this diagram. Seyferts  span a large portion of the diagram, suggesting their coronae can be classified according to this diagram while the radio loud objects are concentrated in the same region, with mean hardness alone a good classifier of the X-ray spectral behavior.
    
    \subsection{The HR track}
\begin{figure}
        \includegraphics[width=\linewidth]{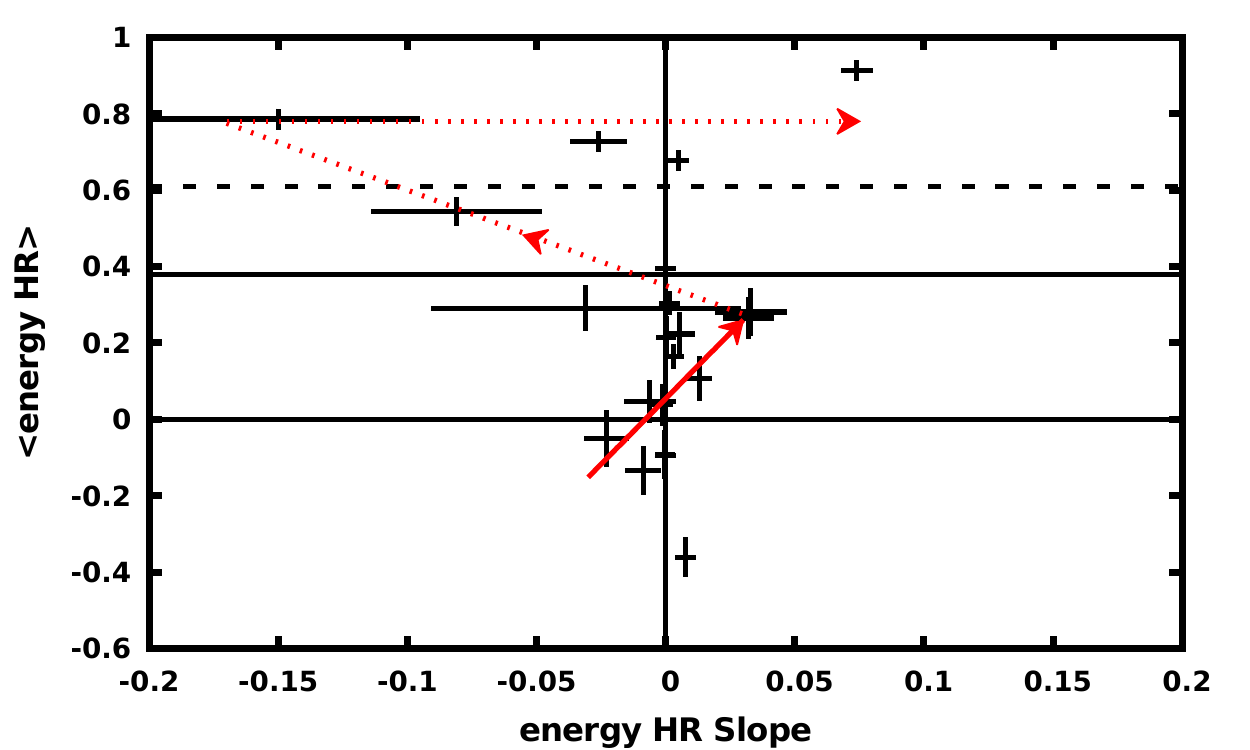}
        \caption{A possible track for Seyferts on the HR-HR-slope (energy based) diagram. The solid arrow is a best fit regression line of slope 6.7, 
            and the dashed lines continue one proposed continuation of the track, just to guide the eye.}\label{fig:track}
    \end{figure}

    In this section we will consider only HR in terms of energy. Looking at Fig. \ref{fig:phase_diagrams}(d), since the radio loud objects seem to be classified completely by their mean HR it makes sense to focus separately on the Seyferts. The sample is small, but there is a possible track in the phase diagram, as shown in Fig. \ref{fig:track}.

    In this figure the first, lower branch is fit with a simple linear regression 
    yielding a slope of 6.7, $R^2=0.73$, and a p-value of 0.007. 
    Akn~564, the Seyfert at the bottom right, is much softer than the rest of the Seyfert population, and is excluded from this fit.
    The track may continue in multiple ways into the absorbed AGN region of the track in the top left. One option is shown by the dashed lines, which are not fits.
    NGC 3227 is an outlier with an energy HR-slope of -0.7, not plotted in Fig.~\ref{fig:track}, and thus the track could be further skewed to the left. In this proposed track the objects become harder-when-dimmer with increasing hardness, until saturation and a migration towards the hard part (right hand side) of the plot. 
    
    When considering in particular the first, main branch of the track, Compton cooling dominates its beginning and transitions into coronae dominated by energy injection. Assuming the coronae are similar in nature between Seyferts, such that the heating and cooling mechanisms are ubiquitous, it seems geometry is a simple explanation for the gradual difference. In this case the softer-when-brighter start would be objects with coronae mostly above and around the black hole, a-la the lamppost picture. In this scenario the corona is exposed to remote and cool parts of the disk. 
    As the track trends harder and harder-when-brighter (right and up along this branch), this could be due to the corona dropping towards the black hole. 
    In this part of the branch, coronal heating becomes significant, dominating the reduced cooling.
    Variations in flux will then be mostly due to the energetics of the corona itself and the hardening of the Compton scattering.
    Whether this is a plausible explanation or not needs to be backed up with a dynamical physical model of the corona. 
    
    Beyond the first branch, the few remaining Seyferts track back (to the left, top of the diagram), as obscuration of the AGN becomes important. 
    This could be due to the hard objects, and low corona, giving rise to a significant outflow, coming from the inner disk. 
	This branch reaches an observed saturation at $\left<HR\right> \approx 0.8$, where the soft band is almost completely obscured. 
    The track turns back to the right, maintaining this saturated mean HR, but an increasing harder-when-brighter behavior, up to NGC 4151.

	\subsection{Could there be an AGN cycle?}\label{sec:cycle}
	\begin{figure}
		\includegraphics[trim=0 0 0.25cm 0,clip,width=0.965\linewidth]{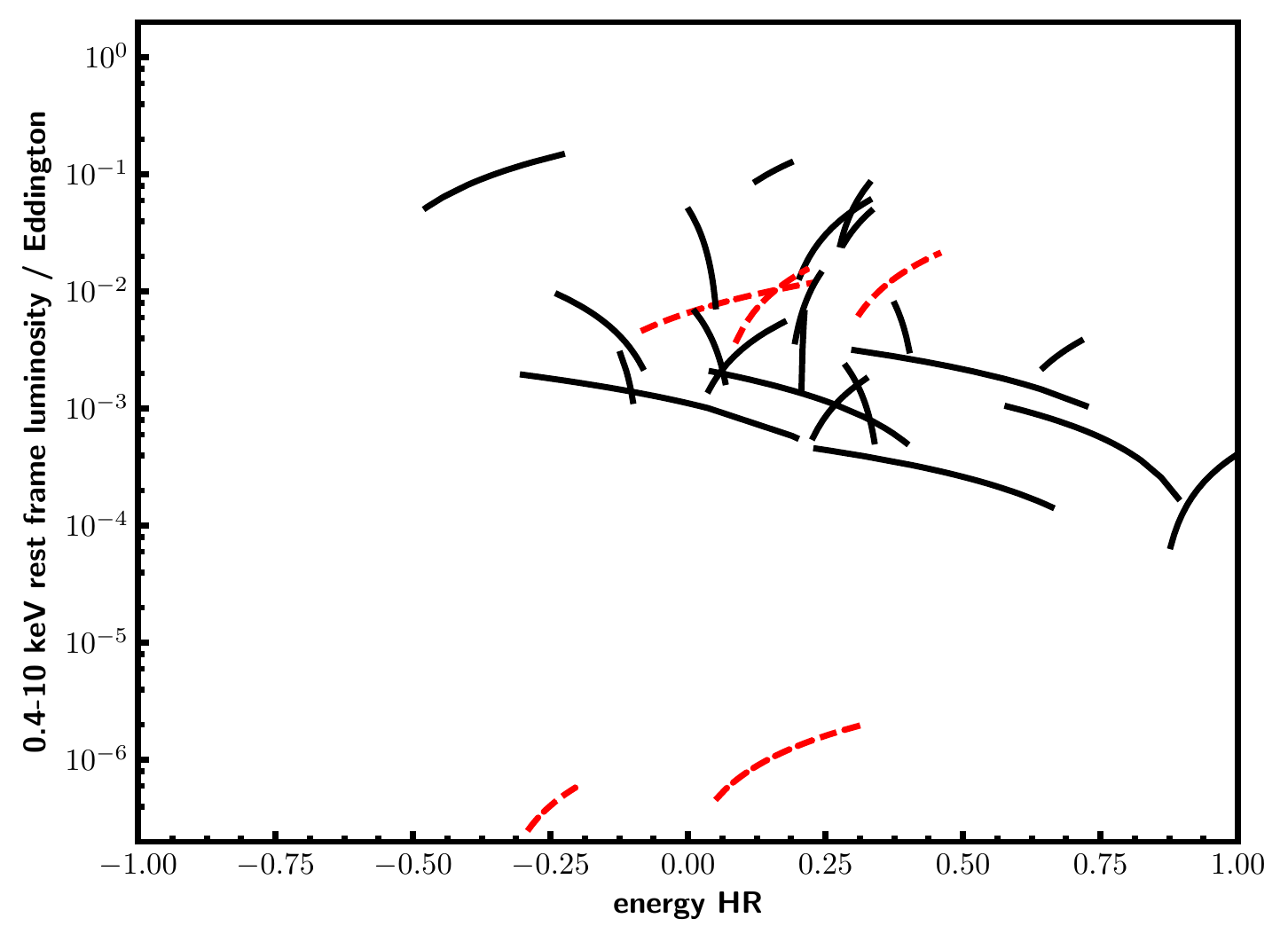}
		\includegraphics[trim=1cm 0 0.25cm 0cm,clip,width=0.96\linewidth]{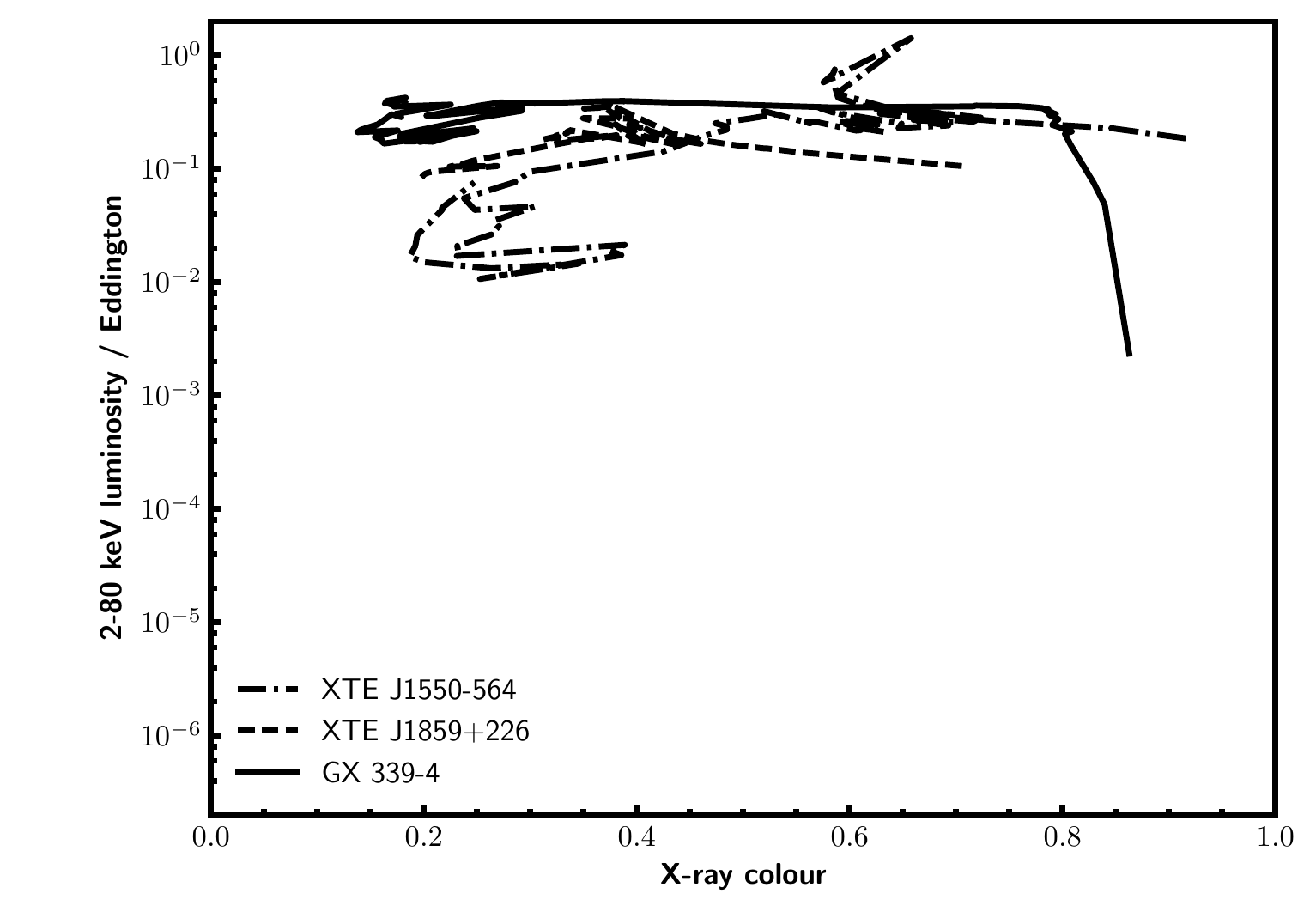}
		\caption{
			$L_X/L_\mathrm{Edd}$  against best fit energy HR (\textbf{top}) for the 27 AGN with available masses (Table \ref{table:sample}). Each line represents the change of HR with $L_X/L_\mathrm{Edd}$ across all available observations.  
            The three upper, red, dashed lines are radio loud Blazars, left to right: 3C273, PKS0548-322, and H1426+428. The lower two are M 87 and M 81, LINERs.
            (\textbf{bottom}) Data of three BHXRBs shown also in Fig. 2 of \citet{Fender05}.
            See Sec. \ref{sec:cycle} for differences of the horizontal axes.}\label{fig:cycle}
	\end{figure}
	There is a possibility that AGN follow a similar cycle to that of BHXRBs. On the other hand, a simple scaling of relevant times such as the viscosity time needed to disperse an accretion ring which scales with the size of the accreting system, $\sim10^6$ years for AGN \citep[e.g.,][]{Duschl00}, means we will not see such a full cycle. 
    Thus, a statistical approach should be used to try and positively discern any such cycles in AGN. 

	Considering that the overall luminosity of an object speaks both to its size and the accretion efficiency, perhaps a better way to compare different objects and their HR would be through $L_X/L_\mathrm{Edd}$.
	We present in the top panel of Fig. \ref{fig:cycle} a plot of the best fit energy HR tracks for the 27 objects that have measured black hole masses (See Table \ref{table:sample}). 
    The three top red, dashed lines are radio loud AGN with mass measurements, 3C273, S5 0716+71, and H1426+428. These radio loud Blazars overlap in this plot with the Seyferts. 
    The red, dashed lines at the bottom are the LINERs M 81 and M 87, separate from the rest of the objects.
	
	Viewed this way there does not seem to be very much of a cycle, as most objects populate the same region of the plot, Seyferts and radio loud AGN. 
	This coexistence on the phase diagram hampers a physical distinction between jetted and non-jetted AGN in terms of HR behavior.
    We compare this plot with BHXRB data from Fig. 2 of \citet{Fender05}, plotted in the bottom panel of Fig. \ref{fig:cycle}.
    Viewed in this way it would seem that all objects presented in Fig. \ref{fig:cycle} are in the hard state extending to the intermediate state as defined by \citet{Fender05}.
    
    Hardness is defined differently in the two panels.    
    We use the definition of HR from equation \ref{eq:hr} with flux calculated from equation \ref{eq:energyflux}.
    \citet{Fender05} define the X-ray color as the counts in the 6.3-10.5 keV band over the counts in the 3.8-6.3 keV band, though different authors may use different bands.
    Since BHXRBs are all thought to follow the same track on this diagram, we assume the full HR axis as we defined it, normalized between -1 and 1 (top panel), should match the total X-ray color axis (bottom panel). Note also the different energy bands used to define $L_X$ in the two panels.
        
    The appearance of M 87 and M 81 in the diagram at low luminosity is interesting as it may provide tentative evidence for AGN in the low state, if there is indeed a cycle. 
    \citet{Markoff15}
    show that the same model, scaled, fits emission in both V404 Cyg and M 81, a stellar mass black hole and a super-massive one, both with similar $L/L_\mathrm{Edd}$. Not many BHXRBs are observed in such a low accretion state, and this analysis implies that a reverse analogy from LINER AGN may give a better understanding to an extremely low accretion state in BHXRBs.
    
    Finally, note  rise and reversal of slope in the luminous top of Fig. \ref{fig:cycle}. This tip has been observed in several BHXRBs, including in XTE J1550-564 shown in the bottom panel.
    GRO J1655-40, which is known for its 2005 flare, displays an even more striking tip \citep[][Fig. 2]{Debnath08}.
	
	Seyferts in Fig. \ref{fig:cycle} mostly occupy the hard part of the diagram as also found by \citet{Connolly16}, though now we address energetics as well. This also strengthens the claim by \citet{Falcke04}, that Seyfert AGN are the hard or intermediate counterpart of BHXRBs. On the other hand, the hard state of BHXRBs has more radio emission (associated with a jet) than the soft state, while Seyferts are known to be radio quiet.
	
	The disk in AGN emits predominantly in the UV, compared to BHXRBs where it emits as soft X-rays. 
	As a consequence, \citet{Kording06} uses simultaneous X-ray and UV observations to estimate HR, or a disk-fraction (UV) - luminosity diagram.
	Their normalized (between 0 and 1) definition captures the hardness state of the entire disk + corona system in a more precise way.
	Nonetheless, they find low luminosity AGN occupy the hard state of the diagram, as in Fig. \ref{fig:cycle}.
	
    \subsection{A short note on other relations}
    In the course of analyzing the hardness behavior of the sample we attempted to find relations of the quantities measured here, $F_\mathrm{var}$, HR, and HR-slope with both $L/L_\mathrm{Edd}$ and radio loudness ($L_R/L_X$). 
    There is only a weak softer-when-brighter trend of mean HR  with $L_X/L_\mathrm{Edd}$ seen only in Seyferts, see Fig. \ref{fig:cycle}. 
    This lack of correlation also holds for the HR slopes. Both of these are true when considering either counts or energy for the HR.
    Finally, none of these quantities show any correlation with X-ray radio loudness.

	\section{Summary and Conclusions}
    
    We analyze 44 AGN observed 20 times and above with \swift XRT, and measure their HR when considering both count and energy definitions. All HR definitions are flux based (cm$^{-2}$ s$^{-1}$) and consistently use the same hard (2-10 keV) and soft (0.4-2 keV) rest frame bands in all AGN. It is important that the soft band encapsulates the soft X-ray excess observed in many AGN, below 2 keV. These definitions are instrument-independent and can be compared with measurements in other X-ray instruments.
    
    We find that using counts HR provides a clear separation of the harder-when-brighter radio loud and the softer-when-brighter radio quiet AGN. The radio loud X-ray variability is dominated by the hard band, in contrast to Seyferts where it is dominated by the soft band, as evident by comparing $F_\mathrm{var}$ in the hard and soft bands. When using energy flux in the HR definition this dichotomy disappears, and radio loud AGN are mixed with Seyferts. Radio loud AGN remain harder-when-brighter or consistent with a flat change of spectrum \citep[see also ][]{Brinkmann03,Ravisio04,Pandey17,Pandey18}, while Seyferts track back and forth across the HR phase diagram (Fig. \ref{fig:phase_diagrams} and \ref{fig:track}).
    
    This energetic analysis implies radio loud and radio quiet AGN should not be discriminated by their HR behavior. This is expected as the  physical origin of the X-ray emission is likely different in the two populations, as the X-ray emission of radio loud AGN is dominated by a jet. 
    This is also seen in \citet{Trichas13} and \citet{Svoboda17}, who through simultaneous observations in UV and X-ray find no true dichotomy of radio loud and radio quiet AGN in the HR diagram (or equivalent), with all objects populating the hard and luminous part of the diagram.  
    We do not have a good physical explanation for the dichotomy observed when considering count HR (Upper left panel of Fig. \ref{fig:phase_diagrams}).
    
    Considering energetics allows us to attribute physical scenarios to different regions of the HR phase diagram, such that harder-when-brighter objects can be considered objects with active and variable coronae, and soft, softer-when-brighter objects have Compton cooling coronae. The HR behavior trend can be interpreted in terms of the location of the corona in the Seyfert system above the disk plane. While the Seyferts are complex and show diverse behavior, radio loud AGN are completely characterized by their mean HR as might be expected for jet dominated objects.
    
    We attempt to place the 27 AGN with measured black hole mass on a HR-$L_X/L_\mathrm{Edd}$ diagram (Fig. \ref{fig:cycle}), comparing with the similar BHXRB diagram. While all Seyferts populate the hard, luminous state of the diagram, three radio loud AGN are observed inseparably from the Seyferts, somewhat hampering claims that the two can be separate branches of a unified cycle. The only tentative evidence for a cycle are two LINERs, M 81 and M 87, that are observed in the soft, dimmer part of the diagram, and may provide a counterpart to quiescent BHXRBs, if considered with the Seyferts in a unified scheme.
    
    Finally, we suggest a possible track on the HR-HR-slope phase diagram for Seyferts, when defined in energy (Fig. \ref{fig:track}). This track may describe a transition from coronae above their host black hole to coronae which compromise the inner part of a thick disk. 
    
    The present analysis shows that the fast changes on daily and shorter timescales of the flux and spectral shape of the X-ray emitting region of AGN provides a window to the nature of these coronae. Future works need to populate the HR phase diagram with a statistically complete sample of AGN, that is not X-ray selected, and test the hypothesis of a HR track for Seyferts. 
	
	\bibliographystyle{mnras}	
	\bibliography{hr,agn}			
\end{document}